# Magnetoelastic Coupling in Hole-doped Two-dimensional *β*-PbO

*Yan Liang, Xingshuai Lv[*], Thomas Frauenheim[*]*

Yan Liang

Bremen Center for Computational Materials Science, University of Bremen, 28359 Bremen, Germany.

Xingshuai Lv

Shenzhen JL Computational Science and Applied Research Institute, 518109 Shenzhen, P.R. China.

Prof. Thomas Frauenheim

Bremen Center for Computational Materials Science, University of Bremen, 28359 Bremen; German; Shenzhen JL Computational Science and Applied Research Institute, 518109 Shenzhen; P.R. China; Beijing Computational Science Research Center, 100193 Beijing, P.R. China.

E-mail: lvxingshuai@csar.ac.cn (X.L); thomas.frauenheim@bccms.uni-bremen.de (T.F.)

**Abstract**:

The realization of intertwined ferroelasticity and ferromagnetism in two-dimensional (2D) lattices is of great interest for broad nanoscale applications, but still remains a remarkable challenge. Here, we propose an alternative approach to realize the strongly coupled ferromagnetism and ferroelasticity by carrier doping. We demonstrate that prototypical 2D *β*-PbO is dynamically, thermally and mechanically stable. Under hole doping, 2D *β*-PbO possesses ferromagnetism and ferroelasticity simultaneously. Moreover, the robustness of ferromagnetic and ferroelastic orders are doping tunable. In particular, 2D *β*-PbO features in-plane easy magnetization axis that is coupled with lattice direction, enabling ferroelastic manipulation of spin direction. Our work highlights a new direction for 2D magnetoelastic research and enables the possibility for multifunctional devices.

## I. Introduction

Multiferroics refer to an important quantum matter that simultaneously possess more than one ferroic order parameters, namely (anti)ferromagnetism (AFM/FM), ferroelasticity (FEL) and ferroelectricity (FE), has brought great opportunities for applications in spintronics, transistors and data storage [1-4]. In this material class, both multiple ferroic orders and strong coupling between them are extraordinary desirable for nonvolatile modulation one or two ferroic orders by using another. In accompany with the continuing quest for miniaturized electronic devices, the upsurge of two-dimensional (2D) multiferroics is promoted [5-8]. Among all the multiferroics, magnetism based multiferroics stand at the paramount place due to its best aspects of low-cost reading combined with efficient mechanical/electrical writing for nonvolatile memory [9]. However, FE-AFM/FM multiferroics are rarely reported in single layers because of the inherent mutual exclusion between magnetism and FE, dubbed as "$d^0$ rule" [10]. FEL, in contrast, shows better compatibility with AFM/FM because neither symmetry nor partial filled orbitals are specially required, which serves as a fantastic playground for exploring emergent physics at 2D limit. Nevertheless, AFM/FM-FEL systems has been scarcely reported, arising from the fact that 2D magnets are rather rare itself [11-14]. To date, several 2D systems with FEL and AFM/FM are proposed [9,15-20], but strong magnetoelastic effect only exist in a few cases, including FeOOH [21], MnNX [22], CrSX [23], $VF_4$ [24], and $VTe_2$ [17].

The model of carrier doping-induced Stoner ferromagnetism has been well established in many archetypal 2D systems, such as GaSe [25], $α$-SnO [16] and phosphorene [26], which provide a valuable alternative to obtain both ferroelastic and FM orders in 2D single layers. Recently, carrier doping-induced multiferroicity has been predicted in several 2D systems, but none of this category exhibits coupled FEL and FM so far [15,16,27-29]. For example, doping $α$-SnO and PbO single layers indeed succeed in realizing multiferroic phases present FM and FEL simultaneously, but two ferroic order parameters in all of them just coexist due to the robust out-of-plane easy magnetization, excluding any magnetoelastic coupling [15,16]. To this end, realizing strong magnetoelastic coupling in 2D limit, using doping as a start point, is highly desirable for both fundamental scientific interest and technological applications.

In this paper, based on first-principles, we report the discovery of strong magnetoelastic effect in 2D $β$-PbO by hole doping. It is demonstrated that semiconducting 2D $β$-PbO prototype can be easily accessed from the layered bulk counterpart. Moreover, 2D $β$-PbO features strain-driven 90º variant switching with moderate barrier and strong transition signal, suggesting intrinsic FEL. Furthermore, the non-dispersive nature of the valence bands leads to the electronic instabilities, exhibiting both FM and ferroelastic orders simultaneously under hole doping. Most strikingly, hole-doped 2D $β$-PbO shows robust in-plane easy magnetization with tunable magnetic anisotropy that is closely linked with FEL, enabling strong magnetoelastic effect, namely, doping-induced in-plane FM can be preciously manipulated via reversible ferroelastic strain.

## II. Results and Discussions

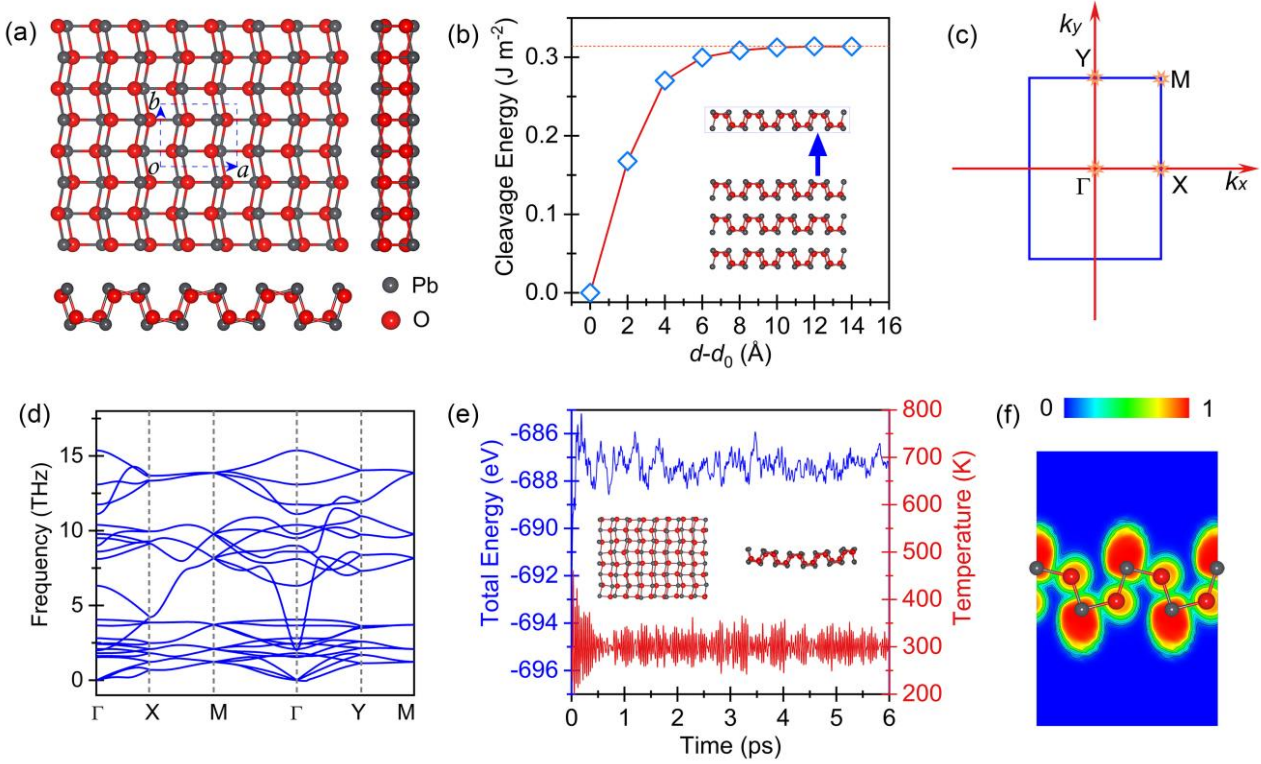

FIG. 1. (a) Top and side views of 2D β-PbO. (b) Variation of total energy with respect to $d-d_0$ for 2D β-PbO, where $d$ is the distance between the top β-PbO layer and the underlying layers. $d_0$ is the equilibrium distance. The inset shows the exfoliation process. (c) 2D Brillouin zone. (d) Phonon spectrum for 2D β-PbO. (e) Top and side views and total energy of β-PbO supercell captured from the end of AIMD simulations at 300 K and the corresponding fluctuation of total energy and temperature. (f) ELF for 2D β-PbO.

The crystal structure of bulk β-PbO is shown in Figure S1 in the Supplemental Material, which is crystallized in yellow orthorhombic form with van der Waals layered structure and has been synthesized several decades ago [30-32]. The atomic layer has a tetragonal symmetry and consists four Pb and four O atoms per unit cell, in which Pb (O) is four-coordinated with adjacent O (Pb) atoms as presented in Fig. 1(a) for top and side views of 2D β-PbO. The lattice constants of 2D β-PbO are found to be $a = 5.80$ Å and $b = 4.76$ Å, which is quite close to its bulk phase [32]. The weak interlayer van der Waals forces in bulk β-PbO suggests that monolayer β-PbO could be exfoliated from its layered bulk. We evaluate the possibility of mechanical or liquid isolation by calculating the cleavage energy. The resulting cleavage energy as a function of separation $d-d_0$ is shown in Fig. 1(b), the total energy increases gradually with the separation distance and finally converges to a constant value of ~ 0.31 J m$^{-2}$. This value is smaller than that of graphene (0.37 ± 0.01 J m$^{-2}$) [33], Nb$_2$ATe$_4$ (0.42/0.43 J m$^{-2}$) [34] and GeS (0.52 J m$^{-2}$) [35], which firmly indicates that the experimentally exfoliation of 2D β-PbO from its bulk is highly feasible.

To examine the stability of 2D β-PbO, the phonon spectrum is first calculated as depicted in Fig. 1(d). No imaginary phonon frequencies are observed in the entire Brillouin zone, confirming the dynamical stability of 2D β-PbO. The thermal stability is further checked by performing AIMD simulations. After heating at 300 K for 6 ps, the total energy of 2D β-PbO only fluctuate slightly and no broken bonds or structure reconstruction appear [Fig. 1(e)], demonstrating that 2D β-PbO is rather stable to keep their structural integrity at room temperature. Additionally, the four independent elastic constants are calculated to be $C_{11}$ = 20.03 N/m, $C_{22}$ = 59.98 N/m, $C_{12}$ = 16.18 N/m and $C_{66}$ = 20.74 N/m, respectively, which meet the Born criteria for mechanical stable 2D lattice: $C_{11}C_{22} - C_{12}^2 > 0$ and $C_{66} > 0$ [36]. We also characterize the bonding types in 2D β-PbO by calculating its electron localization function (ELF), which is plotted in Fig. 1(f). Substantial concentrations of electrons are distributed at the Pb and O sites as well as the centers between them, reflecting a clear covalent bonding feature.

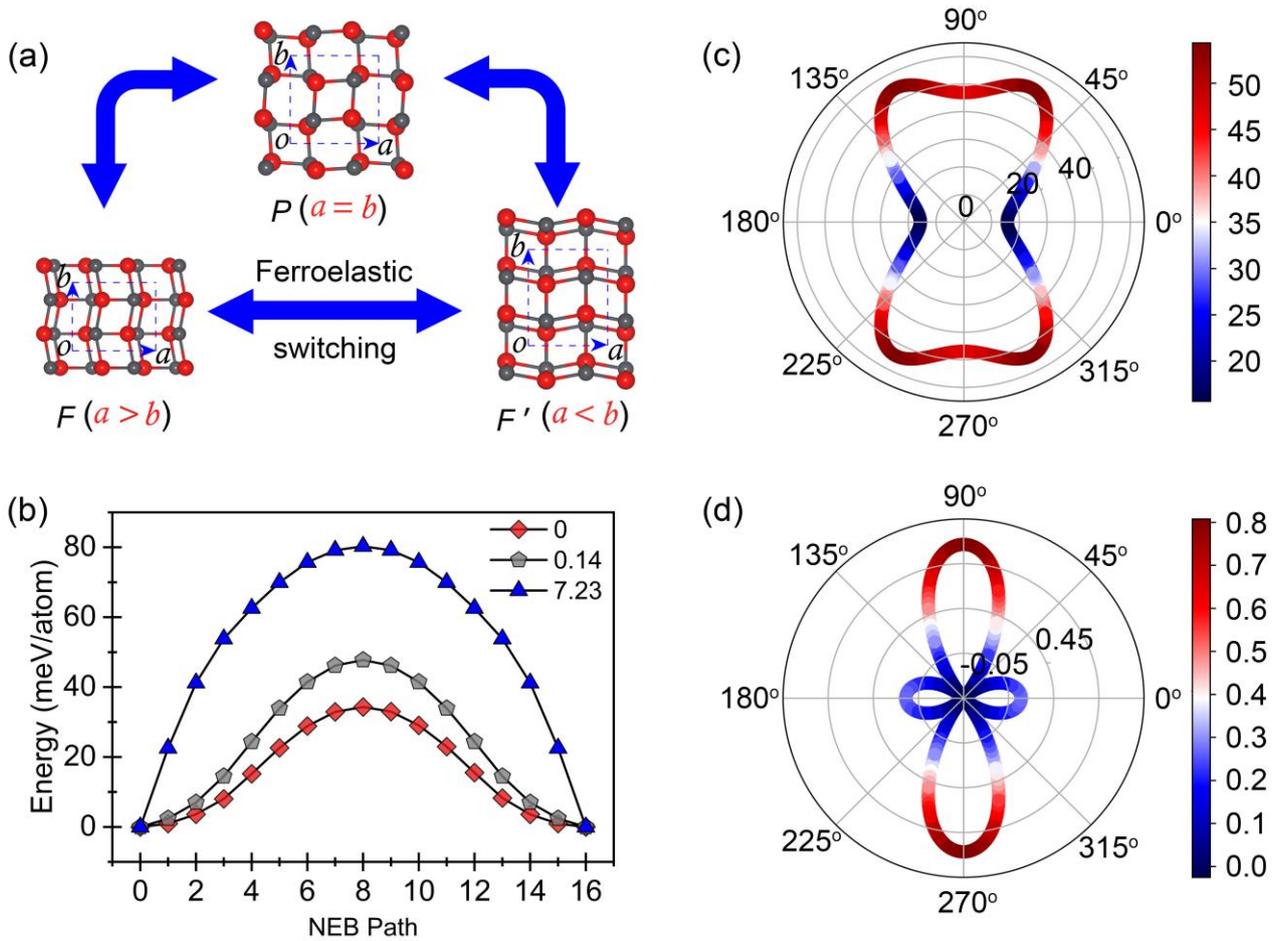

FIG. 2. (a) Schematic plot of ferroelastic switching path between two ferroelastic states $F$ and $F'$ for 2D β-PbO. (b) Energy barriers of ferroelastic switching under hole doping concentration of 0 × $10^{14}$/cm², 0.14 × $10^{14}$/cm² and 7.23 × $10^{14}$/cm². The direction-dependent (c)Young's modulus and (d) Possion's ratio of 2D β-PbO.

FEL is defined by the coexistence of multiple orientation variants of spontaneous lattice strain, which can be switched from one variant to another energetically equivalent state under external mechanical stimuli. Hence, FEL is expected in 2D $\beta$-PbO taking the crystal symmetry of 2D $\beta$-PbO into account. The two orientation states with the same energy (states $F$ and $F'$) are presented in Fig. 2(a). For ferroelastic state $F$, lattice parameter $a$ is larger than $b$. Under the external compressive strain along $a$ axis, the larger lattice switches to $b$ axis. That is, ferroelastic state $F$ transforms to another ferroelastic state $F'$, in which $|a'| = |b|$ and $|b'| = |a|$. As a result, the lattice transformation trace between $F$ and $F'$ can be regarded as the result of 90° rotation operation on $F$ or $F'$ state. The intermediate state, also refer to paraelastic state $P$, that connect the transition pathway is a centrosymmetric square lattice with lattice constants $|a| = |b| = 5.29$ Å. As displayed in Figure S2, the phonon spectrum of paraelastic state $P$ shows imaginary modes around $\Gamma$ and M points, suggesting it is unstable. Thus, the relaxation of paraelastic state $P$ is spontaneous, giving rise to ferroelastic states $F$ or $F'$.

To get deeper understanding on the FEL of 2D $\beta$-PbO, the transformation strain matrices $\eta$ for the ferroelastic transition is studied based on the ferroelastic states $F$, $F'$ and paraelastic state $P$. Transformation matrix $\boldsymbol{J}$ relates to $F$, $F'$ and $P$, and the transformation strain matrix $\eta$ can be obtained by the definition of Green-Lagrange strain tensor [37]:

$$\eta = \frac{1}{2}(\boldsymbol{J}^T\boldsymbol{J} - \boldsymbol{I}) = \frac{1}{2}([\boldsymbol{H}_P^{-1}]^T \boldsymbol{H}_I^T \boldsymbol{H}_I \boldsymbol{H}_P^{-1} - \boldsymbol{I})$$

where superscript $^T$ is the matrix transpose, $\boldsymbol{I}$ is a 2 × 2 identity matrix. $\boldsymbol{H}_P = \begin{vmatrix} 5.29 & 0 \\ 0 & 5.29 \end{vmatrix}$ and $\boldsymbol{H}_I = \begin{vmatrix} 5.80 & 0 \\ 0 & 4.76 \end{vmatrix}$ are the lattice constant matrices for the paraelastic state $P$ and ferroelastic states $F$, respectively. Then, the strain matrix $\eta = \begin{vmatrix} 0.12 & 0 \\ 0 & -0.10 \end{vmatrix}$ is obtained, implying that there is 10% compressive strain along $b$ direction and 12% tensile strain along $a$ direction for ferroelastic state $F$, in comparison with paraelastic state.

In order to judge the excellence (feasibility and robustness) of FEL in 2D $\beta$-PbO, the energy barrier for ferroelastic transition is first calculated using NEB method. As shown in Fig. 2(b), in view of structural symmetry, the revolution paths of $P \to F$ and $P \to F'$ are identical. The intrinsic energy barrier for 2D $\beta$-PbO is obtained to be 34.32 meV/atom. This value is higher than those of SnS (4.2 meV/atom) [7] and ScP (2.53 meV/atom) [39] but much lower than those of GaTeCl (0.16 eV/atom) [39] and CrSX (X = Cl, Br, I) (116 - 178 meV/atom) [23]. This moderate energy berrier not only facilitates the ferroelastic orientation switching under mechanical stimuli, but also highlights the robustness of ferroelastic states against external perturbation. Another criterion that determines the ferroelastic performance is the reversible ferroelastic strain, which is defined as [(|a/b| - 1) × 100%]. The high ferroelastic strain for 2D $\beta$-PbO reaches 22%, comparable or larger than those of frequently reported ferroelastics [9,22,34,40], ensuring the distinct ferroelastic signal intensity of 2D $\beta$-PbO that in favor of shape memory applications.

Actually, structural rigidity and transverse synthetic strain of the system against corresponding axial strain play decisive roles in FEL. High flexibility and positive Posson's ratio along ferroelactic variant orientations are two indispensable preconditions. Thus, Young's modulus $Y(\theta)$ and Poisson's ratio $v(\theta)$ of 2D $\beta$-PbO are investigated based on the elastic constants above. Young's modulus $Y(\theta)$ and Poisson's ratio $v(\theta)$ can be obtained by the expression as follows [41]:

$$Y(\theta) = \frac{C_{11}C_{22}-C_{12}^2}{C_{22}\cos^4\theta+C\cos^2\theta\sin^2\theta+C_{11}\sin^4\theta},$$

$$v(\theta) = \frac{C_{12}\cos^4\theta-B\cos^2\theta\sin^2\theta+C_{12}\sin^4\theta}{C_{22}\cos^4\theta+A\cos^2\theta\sin^2\theta+C_{11}\sin^4\theta}.$$

Here, $A = (C_{11}C_{22} - C_{12}^2)/C_{66} - 2C_{12}$ and $B = C_{11} + C_{22} - (C_{11}C_{22} - C_{12}^2)/C_{66}$. $\theta = 0°$ corresponds to $a$ axis. The $\theta$-dependent values of $Y(\theta)$ and $v(\theta)$ are presented in Fig. 2(c) and 2(d). Young's modulus of 2D $\beta$-PbO exhibits strong anisotropy due to the anisotropic structural morphology. Besides, values of Young's modulus for 2D $\beta$-PbO range from 15.66 - 54.17 N/m, which is comparable with AgF$_2$ (32 N/m) [9], 1$S'$-MSSe (M = Mo, W) (70-72 N/m) [42] and GaTeCl (42 N/m) [39] and much smaller than those of $\alpha$-MPI (141-152 N/m) [43] and graphene (340 N/m) [44], indicative of the superior mechanical flexibility and thus it is easier to apply strain in the in-plane directions for switching the FEL. As for Poisson's ratio, it also shows anisotropic behavior. Poisson's ratio is obtained to be 0.27 (0.81) along $a$ ($b$) axis, suggesting sensitive structural deformation under external stress. And interestingly, negative Poisson's ratio of -0.025 is found along diagonal direction in 2D $\beta$-PbO, as depicted in Figure S3.

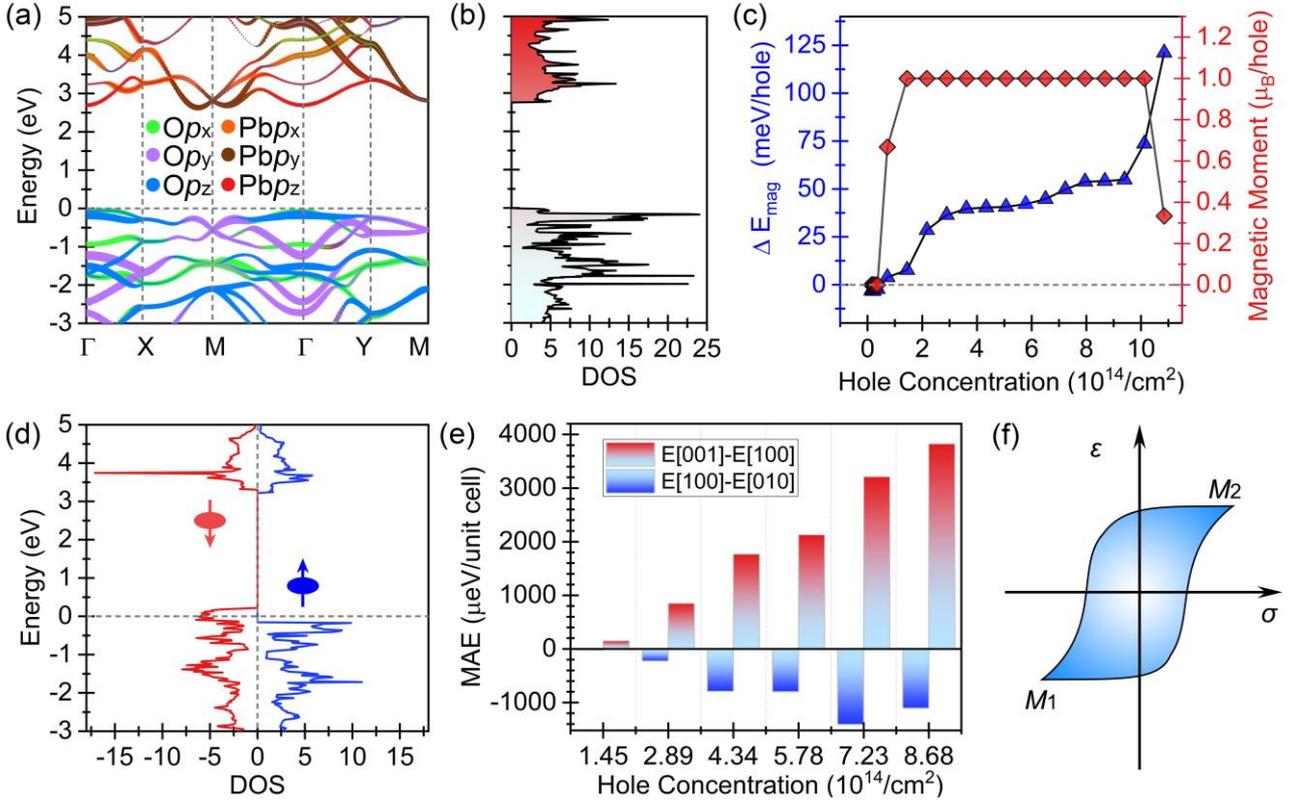

FIG. 3. (a) Orbital resolved band structure and (b) density of states of 2D β-PbO. (c) Variation of the magnetic energy and magnetic moment (per carrier) as a function of doping concentration. (d) Density of states under the hole doping concentration of $3.61 \times 10^{14}/cm^2$. (e) Doping-dependent in out-of-plane and in-plane MAEs. (f) Schematic representation of ferroelastic hysteresis loop that is coupled with doping-induced magnetism in 2D β-PbO. Two magnetic states $M_1$ and $M_2$ possess orthogonal easy magnetization axes.

We next focus on the electronic properties of 2D β-PbO. The orbital-resolved band structure along high-symmetric points is shown in Fig. 3(a). Clearly, 2D β-PbO is an indirect band gap semiconductor, with the valence band maximum (VBM) locates along Γ-X line and conduction band minimum (CBM) sits between X-M. The band gap is calculated to be 2.61 eV at the PBE functional, whereas HSE06 correction to the PBE gap is about 1.10 eV. In addition, CBM of 2D β-PbO is dominated primarily by $p_y$ and $p_z$ prbitals of Pb, while VBM is mainly formed by $p_y$ and $p_z$ orbitals of O. Another interesting point of the band structure worth noting is that valence bands are distinctive by very flat dispersion in the momentum space. This is reasonable as valence bands near the Fermi level is O-dominated, while the bonding environment results in the rather weak p-orbital coupling between the nearest neighboring O atoms and is responsible for the less dispersed valence bands. As illustrated in Fig. 3(b), The flat valence bands create multiple saddle points in the vicinity of X, M and Y points, leading to a remarkably extended van Hove singularity in the density of states (DOS) that is first peaked at the energies of 0.17 eV below the VBM. According to band-picture model, spontaneous ferromagnetic ground state emerges when relative exchange interaction is larger than the loss in kinetic energy, that is, Stoner criterion $N(E_F)I > 1$ [45]. Therefore, Stoner ferromagnetism is expected in 2D $\beta$-PbO, once doping pushes Fermi level approaching the divergent points.

To verify the electronic instability of 2D $\beta$-PbO driven by dole doping, Fig. 3(c) shows calculated concentration-dependent magnetic energy per carrier, $\Delta E_{mag}$ (i.e., total energy difference between non-magnetic phase and ferromagnetic phases), and magnetic moment (i.e., $\sum_{n\mathbf{k}\in hole} -<n\mathbf{k}|\mathbf{m}|n\mathbf{k}> / \sum_{n\mathbf{k}\in hole} <n\mathbf{k}|n\mathbf{k}>$, where $\mathbf{m}$ and $|n\mathbf{k}>$ denote spin magnetic moment operator and the Bloch states). Ferromagnetic state begins to be more stable when the hole concentration is larger than $0.72 \times 10^{14}/cm^2$ (0.2 hole per unit cell). Then, larger spin magnetic moment develops as concentration increase, and saturates at 1 $\mu_B$/hole upon hole density between 1.45 - 10.12 $\times 10^{14}/cm^2$ (0.4 - 2.8 hole per unit cell). Further increment of doping concentration reduces the magnetic moment. In contrast to magnetic moment, magnetic energy $\Delta E_{mag}$ increases monotonically as a function of doping concentration. In Fig. 3(d), we plot the spin-polarized DOS of 2D $\beta$-PbO under the doping level of $3.61 \times 10^{14}/cm^2$ (1 hole per unit cell). Compared with pristine 2D $\beta$-PbO, doping causes a considerable energy splitting of 0.49 eV between spin-up and spin-down channels with Fermi level traverse only spin-down bands, indicating its fascinating prospect for controllable half-metallic spintronics. Given that carriers doping concentration of $\sim 10^{15}/cm^2$ has already been realized experimentally [46,47], high efficiency carrier injection and ferromagnetism in metal shrouded 2D $\beta$-PbO is readily to be achieved.

It is worth noting that obvious structural phase transition from 2D to one-dimensional-like framework of 2D $\beta$-PbO is observed in the doping process when the doping density surpass $9.40 \times 10^{14}/cm^2$ (2.6 hole per unit cell), which is also the critical point for sudden surge of magnetic energy and is similar to previous report [48]. To show this more clearly, the evolution of lattice constants and final structure are shown in Figure S4(a). Besides, such new metastable phase will experience spontaneous collapsing to the original structure after doped carriers are removed. Moreover, FEL maintains under hole doping concentration smaller than $9.40 \times 10^{14}/cm^2$, whereas both the stability of ferroelastic state and ferroelastic signal intensity of 2D $\beta$-PbO could be significantly enhanced as shown in Fig. 2(b) and Figure S4(b).

To get deeper understanding on the doping-induced ferromagnetism, the MAE of 2D $\beta$-PbO under doping is further investigated by involving SOC. Two in-plane ([100] and [010]) and one out-of-plane directions ([001]) are considered. Fig. 3(e) depicts the MAE as a function of doping density, from which we can clearly find the minimum energy spin orientation is [100]. Therefore, it favors in-plane easy magnetization over out-of-plane, and easy magnetization axis for hole doped 2D $\beta$-PbO lies robustly along the [100] direction throughout the hole doping process. More interestingly, it is found that MAEs in both in-plane and out-of-plane are boosted by doping. For example, the MAE of [001] ([010]) direction is 150 (23) $\mu$eV per unit cell larger than [100] direction at low doping level of $1.45 \times 10^{14}/cm^2$ but becomes as high as 3824 (1104) $\mu$eV per unit cell at doping concentration of $8.68 \times 10^{14}/cm^2$, which is sufficient to stabilize the in-plane ferromagnetic ordering.

As discussed above, 2D $\beta$-PbO hold FEL and ferromagnetism simultaneously upon hole doping. Although the FEL and ferromagnetism have different origins, the mutual coupling between them can be expected. The most favorable in-plane easy magnetization axis is strongly coupled to the lattice direction (i.e., along the longer lattice vector), allowing for precise direction tailor of spin orientation by FEL. As schematically plotted in Fig. 3(f), ferroelastic switching rotate the lattice orientation of 2D $\beta$-PbO by 90º, thus the easy magnetization axis will also undergo a 90º rotation. That's to say, ferroelastic transition switches easy magnetization axis from *a* to *b* axis, achieving strong magnetoelastic effect. Different from doped $\alpha$-SnO [16] and PbO [15], as shown in Figure S5, strong magnetoelastic effect maintains even beyond monolayer. To our best knowledge, this is the first example that systematically demonstrates strong magnetoelastic coupling in 2D limit by carrier doping, and we believe this work will ignite more efforts to forge 2D functionalities via carrier doping.

### III. Conclusion

In summary, we show that stable 2D $\beta$-PbO harbors ferroelastic and ferromagnetic orders simultaneously, depending on the concentration of hole doping. Meanwhile, the robustness of FEL and ferromagnetism are doping tunable. Remarkably, strong magnetoelastic effect is observed in 2D $\beta$-PbO. Under hole doping, easy magnetization axis along the longer lattice vector is closely linked with FEL, rendering ferroelastic control of MAE. These findings imply the possibility of achieving strong magnetoelastic effect in 2D lattices by doping, which are useful for both fundamental research and technological applications.

### Experimental Section

First-principles calculations are performed on the basis of density functional theory as implemented in the Vienna *Ab Initio* Simulation Package (VASP) [49]. The Perdew–Burke–Ernzerhof (PBE) parametrization of the generalized gradient approximation (GGA) is used to account for electron exchange-correlation [50]. Monkhorst–Pack $k$-point meshes [51] of $9 \times 9 \times 1$ and $11 \times 11 \times 1$ are used for geometry optimization and self-consistent electronic structure calculations. The convergence criteria of energy and Hellmann-Feynman force are $1.0 \times 10^{-5}$ eV per atom and 0.01 eV Å$^{-1}$, respectively, at the cutoff energy of 500 eV. The thickness of vacuum space is set to be larger than 18 Å in the direction normal to the basal plane to avoid the interaction between its periodic adjacent layers. Spin-orbital coupling (SOC) is considered in magnetocrystalline anisotropy energy (MAE) calculations. Phonon spectrum is obtained using the finite displacement method as implemented in the Phonopy code [52]. *Ab Initio* molecular dynamic (AIMD) simulation is performed with the NVT ensemble and the temperature is controlled by using the Nosé–Hoover thermostat [53]. The nudged elastic band (NEB) method [54] is adopted to estimate the energy barriers of ferroelastic switching.

**Supporting Information**

Supporting Information is available from the Wiley Online Library or from the author.

**Acknowledgements**

This work is supported by China Postdoctoral Science Foundation under Grant No. 2021M692213, DFG Research and Training Group RTG 2247 "Quantum Mechanical Materials Modelling".

**Conflict of Interest**

The authors declare no conflict of interest